\documentclass[12pt]{article}
\begin{document}
\title{The Berry phase in frustrated spin glass}
\author{Dipti Banerjee$^*$\\
Department of Physics,Rishi Bankim Chandra College \\
Naihati,$24$-Parganas(N),Pin-$743 165$, West Bengal, INDIA \\ and \\
{\bf The Abdus Salam International Center for Theoretical
Physics},\\ Trieste, ITALY}
\date{10.09.07}
\maketitle

\begin{abstract}
In this letter we have pointed out that frustration in spin glass
is realized through the Berry phase due to the conflict between
the spin ordering in the course of parallel transport of spinor.
 We have came to the point that the Berry phase depicting the chiral change
of helicity of a quantized spinor is prominent only in the
presence of frustration.
\end{abstract}

\vspace{3cm}PACS No:-$71.55$.Jv

 \vspace{5cm}

 *email:deepbancu@homail.com;dbanerje@ictp.it

\pagebreak
 In the theory of spin glasses, the concept of frozen
spin configuration and frustration have played an important role
[1]. The frozen spin gives the emphasis on the rigidity opposed to
the spatial ordering of spin. Interaction between the spins are in
conflict with each other due to some quenched disorder leading to
frustration. It is known that various classes of randomness exist
for the behavior of spin glass and this randomness leads to
frustration [2]. From the topological point of view, constraints
prevent the neighboring spins to have the minimum bond energy in
spin glass. The geometry of the spin-ordering in a spin glass has
a similarity with the "parallel transport" of a tangent vector on
a curved surface [3]. The misfit between the various lines of
transport can be expressed by the frustration and curvature
respectively. In this sense frustrated plaquette are curved
whereas un-frustrated are flat. The above ideas imply that 'the
frustration in spin glass' can be realized through the curvature
of space measured by the Berry phase[4]. This topological phase is
developed  by the parallel transport of the spinor over a closed
path. The analogy at the deeper level lies in gauge symmetries
where gauge potentials known as Berry connections are the source
of curvature of space time. The spin Berry phase plays an
important role in the quantum transport of strongly correlated
spin system [5]. This phase is also the very cause of net change
of spin chirality visualized through chiral anomaly in the field
theoretic aspect [6]. In this paper we will focus our attention to
study the frustration of quantum spin glass from the view point of
Berry phase.

In continuous rotations of spins the Hamiltonian of granular spin
glass system is [7],
\begin{equation}
H= -J_{ij}\Sigma_{ij} cos(\phi_i -\phi_j)
\end{equation}
where $J_{ij}$ is a coupling depending on the nature of host
(metal, insulator and superconductor etc.) material. Here the
complex energy gap of ith grain is
$$\psi_i = \Delta_i exp(i\phi_i)$$
This is similar to the Hamiltonian of a XY spin ferromagnet. Due
to short ranged interaction the energy of all domain walls becomes
[1]
\begin{equation}
\Delta E(C) = \Sigma_{ij}J_{ij} cos(\theta_i -\theta _j)
\end{equation}
where $(\theta_i -\theta _j)$ is the angle between the two spins
at ith and jth site respectively. In absence of frustration the
spinors being strongly correlated and $cos(\theta_i -\theta_j)$ is
almost decided by $J_{ij}$. The presence of frustration results
weak correlation determining $(\theta_i -\theta _j)$ not only by
$J_{ij}$ but also by the rest of the neighbors.

The above Hamiltonian changes in presence of magnetic field
\begin{equation}
H= -J_{ij}\Sigma_{ij} cos(\phi_i -\phi_j - A_{ij})
\end{equation}
where
\begin{equation}
A_{ij}=\frac{2\pi}{\Phi_0}{\int_i}^j \vec{A}.\vec{dl}
\end{equation}
is the gauge potential generated by the interaction of two
spinors. Here $\Phi_0$ is the elementary flux quantum $hc/2e$.
This shows that a magnetic field can act as a source of
frustration. Replacing the element of the above spin vectors by
Pauli matrices, we get a quantized model. The above Hamiltonian in
eq.(3) becomes [7]
\begin{equation}
H=J \Sigma({S_i}^* U_{ij} S_j+ h.c)
\end{equation}
where $U_{ij}=exp(iA_{ij})$ represent the link gauge degree of
freedom and $S_i=\exp(i\phi_i)$ the spin vector respectively. Here
the randomness arises from the difference angle $A_{ij}$ rather
than the exchange bonds.
 The Hamiltonian remains invariant under the local
gauge transformation.
$$
{S_i}'\rightarrow V_i S_i$$~~~~~~and $$~~~~~~{U_{ij}}' \rightarrow
V_i U_{ij}{V_j}^*
$$
 where $V_i=exp(i\theta_i)$. In fact under local gauge
transformation applied at the ith site the corresponding spin gets
rotated by an angle $\theta_i$ and each of the connecting links
get rotated by the difference of angle $(\theta_i - \theta_j)$
such that the Hamiltonian remains invariant under such
transformation. For the conventional XY model the matrix
$U_{ij}=\pm 1$ restricting the transformation angles $\phi_i$ to
$(0,\pi)$.

 In a frustrated spin system the relative orientations of neighboring
spins are not only decided by their interaction alone but also by
the rest of the spin society. For any closed path in a lattice
spin the sign of product of the exchange integral is known as
frustrated function. Here the angles $\psi_{ij}$ are treated as
continuous variables, which correspond to complex bonds $J_{ij}$.
In a frustrated system the exchange integral around any closed
contour is equal to $-1$ whereas for the un-frustrated system it
is $+1$. The quantity
\begin{equation}
\exp[2\pi i \phi_{ijkl}] = U_{ij} U_{jk} U_{kl} U_{li}
\end{equation}
is called the frustration function defined for any closed path in
the lattice spins [7]. Considering $U_{ij}=\exp{iA_{ij}}$ the
above equation changes to
$$\phi_{ijkl}=A_{ij}+A_{jk}+A_{kl}+A_{li}$$
This is the case for the frustrated square plateaus. If we
consider the triangular frustrated lattice then the frustrated
function will be
\begin{equation}
\phi_{ijk}=A_{ij}+A_{jk}+A_{ki}=\sum_{ij} A_{ij}
\end{equation}
It seems that this sum over link gauge degree of freedom
$\sum_{ij} A_{ij}$ gives rise in the continuum limit the
connection over a closed path.This implies that a frustrated
function is equivalent to the net change of curvature over the
closed path measured through Berry phase.

A frustrated system is described by a chiral spin liquid where the
signature of chiral spinor $\psi_L$ or $\psi_R$ may be considered
to represent the order parameter. For a system with an odd number
of anti ferromagnetic links this change in chirality of the above
two spinors will lead to a change in chirality in a frustrated
loop [8]. The order parameter of a frustrated spin system can be
depicted by chiral fermions represented by two opposites
orientation of helicities.

It has been pointed out earlier [9] that chiral fermion may be
depicted by a scalar particle moving with $l=1/2$ in an
anisotropic space. In three space dimension, in an axis-symmetric
system where the anisotropy is introduced along a particular
direction, the components of the linear momentum satisfy a
commutation relation of the form
\begin{equation}
[p_i, p_j] = i\mu \varepsilon_{ijk}\frac{x^k}{r^3}
\end{equation}
 Here $\mu$ corresponds to the measure of anisotropy and
behaves like the strength of a magnetic monopole. Indeed the
angular momentum relation in this space is given by
\begin{equation}
\vec{J}=\vec{r}\times\vec{p}- \mu\hat{r}
\end{equation}
with $\mu=0,\pm1/2,\pm1...$. This corresponds to the motion of a
charged particle in the field of a magnetic monopole [10]. The
spherical harmonics incorporating the term $\mu$ becomes
\begin{eqnarray}
{Y_l}^{m,
\mu}=(1+x)^{-(m-\mu)/2}.(1+x)^{-(m+\mu)/2}\nonumber\\
     \times\frac{d^{l-m}}{d^{l-m}x}\left((1+x)^{l-\mu}.
(1-x)^{l+\mu}\right)e^{im\phi}e^{i\mu\chi}
\end{eqnarray}
with $x=cos\theta$.\\
 Since the chirality is associated with the
angle $\chi$ denoting the rotational orientation around the {\it
direction vector} $\xi_\mu$, the variation of the angle $\chi$
i.e. the change of rotational orientation around the direction
vector will correspond to the change in chirality.In spherical
harmonics given by eqn.(10) the spin angular part associated with
the angle $\chi$ is given by $e^{-i\mu\chi}$. Thus when $\chi$ is
changed to $\chi +\delta\chi$, we have
\begin{equation}
i\frac{\partial}{\partial(\chi+\delta\chi)}e^{-i\mu\chi}=
i\frac{\partial}{\partial(\chi+\delta\chi)}e^{-i\mu(\chi+\delta\chi)}e^{i\mu\delta\chi}
\end{equation}
which implies that the wave function will acquire the extra phase
$e^{i\mu\delta\chi}$ due to an infinitesimal change of the angle
$\chi$ to  $\chi+\delta\chi$. When the angle $\chi$ is changed
over the closed path $0\leq\chi\leq2\pi$, for one complete
rotation, the wave function will acquire the phase [10]
\begin{equation}
exp[ i\mu{{\int_0}^{2\pi}}\delta\chi] = e^{2i\pi\mu}
\end{equation}
which represents the spin dependent Berry phase. Indeed in this
formalism, a fermion is depicted as a scalar particle moving in
the field of a magnetic monopole and when a scalar field(particle)
traverses a closed path with one flux quantum ($\mu=1/2$)
enclosed, we have the phase $e^{i\pi}$, suggesting the system a
fermion.

For the specific case of $l=1/2,|m| =|\mu|=1/2$ for half
orbital/spin angular momentum, we can construct from the spherical
harmonics ${Y_l}^{m, \mu}$, the instantaneous eigenstates
$|\uparrow,t \rangle$, representing the two component up-spinor as
\begin{eqnarray}
|\uparrow,t\rangle &=& {u\choose v} = {{Y_{1/2}}^{1/2,1/2} \choose
{Y_{1/2}}^{-1/2,1/2}} \nonumber\\
&=& {\sin\frac{\theta}{2}\exp i(\phi-\chi)/2 \choose
\cos\frac{\theta}{2}\exp-i(\phi+\chi)/2}
\end{eqnarray}
and the charge conjugate state, down-spinor by
\begin{equation}
|\downarrow,t\rangle = {{-Y_{1/2}}^{-1/2,1/2} \choose
{Y_{1/2}}^{-1/2,-1/2}} = {-\cos\frac{\theta}{2}\exp i(\phi+\chi)/2
\choose \sin\frac{\theta}{2}\exp-i(\phi-\chi)/2}
\end{equation}
In an arbitrary superposition of elementary qubits $|0\rangle and
|1\rangle$ the up-spinors becomes
\begin{equation}
|\uparrow,t\rangle =\left(\sin\frac{\theta}{2}e^{i\phi}|0\rangle +
\cos\frac{\theta}{2}|1\rangle\right)e^{-i/2(\phi+\chi)}
\end{equation}

The time evolution of a two state system is governed by an unitary
$SU(2)$ $2\times2$ transformation matrix $U(g)$ as follows
\begin{equation}
U(g) = \pmatrix {{\alpha~~~~-\beta^*}\cr {\beta~~~~~\alpha^*}}
\end{equation}
where $|\alpha|^2 + |\beta|^2=1$ with $|g\rangle=U(g)|0 \rangle $=
$ \alpha\choose \beta$.

These states $|\uparrow,t\rangle$ and $|\downarrow,t\rangle$ can
be generated  by the unitary matrix $U(\theta,\phi,\chi)$
\begin{equation}
U(\theta,\phi,\chi)=
\pmatrix{{\sin\frac{\theta}{2}e^{i/2(\phi-\chi)}~~~~~~~~~
-\cos\frac{\theta}{2}e^{i/2(\phi+\chi)}} \cr
\cos\frac{\theta}{2}e^{-i/2(\phi+\chi)}~~~~~~~~~
{\sin\frac{\theta}{2}e^{-i/2(\phi-\chi)}}}
\end{equation}
from the basic qubits $|0\rangle$ and $|1\rangle$ as follows
\begin{eqnarray}
|\uparrow,t\rangle = U(\theta,\phi,\chi)|0\rangle ,
|\downarrow,t\rangle=U(\theta,\phi,\chi)|1\rangle
\end{eqnarray}

Over a closed path, the single quantized up spinor acquires the
geometrical phase [11]
\begin{eqnarray}
\gamma_{\uparrow}&=& i\oint\langle\uparrow,t|\nabla|\uparrow,t\rangle.d\lambda\\
&=& i\oint\langle0|U\dag dU|0\rangle.d\lambda\\
 &=& i\oint A_{\uparrow}(\lambda)d\lambda\\
 &=&i \oint L^\uparrow_{eff} dt\\
 &=& i\frac{1}{2}(\oint d\chi-\cos\theta\oint d\phi)\\
 &=&i \pi(1 - \cos\theta)
\end{eqnarray}
This shows that for quantized spinor, the Berry Phase is a solid
angle subtended about the quantization axis.
 For conjugate state, the down spinor becomes
\begin{equation}
|\downarrow(t)\rangle = (-\cos\frac{\theta}{2}|0\rangle +
\sin\frac{\theta}{2}e^{-i\phi}|1\rangle)e^{i/2(\phi+\chi)}
\end{equation}
giving rise in similar manner the Berry phase over the closed path
\begin{equation}
\gamma_{\downarrow} = -i\pi(1-\cos\theta)
\end{equation}
 The fermionic or the antifermionic
nature of the two spinors (up/down) can be identified by the
maximum value of topological phase $\gamma_{\uparrow/
\downarrow}=\pm \pi$ at an angle $\theta=\pi/2$.  For $\theta=0$
we get the minimum value of $\gamma_{\uparrow}=0$ and at
$\theta=\pi$ no extra effect of phase is realized.

 This Berry phase visualized by the solid angle is acquired by the parallel
transport of the quantized spinor over a closed path resulting the
reunion of the final point with the initial in the absence of
local frustration. The Berry phase in connection with chirality as
in eq.(12) is not visible here. One can verify that this Berry
phase in eq.(24) remains same if we neglect the overall phase
$e^{-i(\phi+\chi)/2}$ from the quantized spinors in eq.(15). This
is possible when there is no local frustration causing any spin
conflict in the system.

In case of spin glass, the frustrated spinor acquire different
Berry phase. Due to nontrivial frustration by the disorder in the
glassy system, the quantized spinor does not reach the initial
point. In other words the path traced out by the spinor is not
closed. Intuitively the initial and final points are connected by
the fibre representing the gauge due to randomness in the spin
direction.

 Murakami et.al. [5] pointed out that when an electron hops from
site i to j coupled to a spin at each site then the spin wave
function is effectively
\begin{equation}
|\chi_i\rangle=t\left(e^{ib_i}\cos\frac{\theta_i}{2},
e^{i(b_i+\phi_i)}\sin\frac{\theta_i}{2}\right)
\end{equation}
The overall phase $b_i$ corresponding to the gauge degree of
freedom does not appear as physical quantities. The effective
transfer integral $t_{ij}$ is given by
\begin{eqnarray}
t_{ij}&=&t\langle\chi_i|\chi_j\rangle\nonumber\\
      &=&te^{(b_j-b_i)}\left(\cos\frac{\theta_i}{2}\cos\frac{\theta_j}{2}+
e^{i(\phi_j-\phi_i)}\sin\frac{\theta_i}{2}\sin\frac{\theta_j}{2}\right)\nonumber\\
      &=&t e^{i a_{ij}}\cos\frac{\theta_{ij}}{2}
\end{eqnarray}
where $\theta_{ij}$ is the angle between the two spins $\vec{S_i}$
and $\vec{S_j}$. The phase $a_{ij}$ is the vector potential
generated by the spin,and corresponds to the Berry phase felt by
the hopping electron. It has been pointed out [5], that the total
phase obtained by an electron hopping along a loop
$1\longrightarrow 2 \longrightarrow 3 \longrightarrow 1$ is also
the solid angle subtended by the three spins. This $a_{ij}$
measures the spin chirality in the context of quantum spin liquid
where the spins fluctuate quantum mechanically [12].

In the light of above works we concentrate in finding the Berry
phase of a quantized spinor residing on a spherical frustrated
surface. Rotation takes place once through points having the same
solid angle in terms of $\theta$. The transfer integral of the
quantized spinor as in eq.(15) will be
\begin{eqnarray}
\langle{\uparrow,t}_j|{\uparrow,t}_i\rangle =
e^{i/2(\phi_j-\phi_i)}.
e^{i/2(\chi_j-\chi_i)}\nonumber\\
\left(\cos\frac{\theta_i}{2}\cos\frac{\theta_j}{2}+
e^{i(\phi_i-\phi_j)}\sin\frac{\theta_i}{2}\sin\frac{\theta_j}{2}\right)
\end{eqnarray}
Comparing with eq.(28) we have the transfer integral $t_{ij}$
expressing the variation of $\chi$ along with the angle
$\theta_{ij}$.
\begin{equation}
\langle{\uparrow,t}_j|{\uparrow,t}_i\rangle =
e^{i/2(\phi_j-\phi_i)}e^{i/2(\chi_j-\chi_i)}
\cos\frac{\theta_{ij}}{2}
\end{equation}
Representing the change of helicity $(\chi_i-\chi_j)
=\chi+\delta\chi - \chi =\delta\chi$ the corresponding phase
$$e^{i/2(\chi_j-\chi_i)}=e^{i/2 \delta\chi}=e^{i/2 a_{ij}}$$
is the Berry phase (eq.12) visualize the chiral change of helicity
of quantized fermion. The following phase
\begin{equation}
\langle{\uparrow,t}_j|{\uparrow,t}_i\rangle = e^{i/2(\phi_j-\phi_i
+a_{ij})} \cos\frac{\theta_{ij}}{2}
\end{equation}
represents the difference of inclination of helicity over the
virtual closed path. Following the local gauge transformation
\begin{equation}
a_{ij}\longrightarrow a_{ij}+\phi_i-\phi_j
 \end{equation}
 the two eqs.(28) and (31) are equivalent.
 Hence we have a similar form of transfer
 integral for quantized spinor as in [5].

 In a frustrated system the quantized spinors fix up their
helicity. Transportation around a closed loop represents only the
variation of $\phi$ values from $0\rightarrow 2\pi$ where the
slight shift of $\chi$ values is visualized as chiral gauge due to
some conflicts between the spins offered by the disorders in the
system. The required Berry connection of a quantized up spinor in
the frustrated spin system can be obtained after few mathematical
steps using eq.(15)
\begin{eqnarray}
\langle\uparrow_j|d|\uparrow_i\rangle &=&e^{i/2(\phi_j-\phi_i)}.
e^{i/2(\chi_j-\chi_i)}.\nonumber\\
&&\left(\sin\frac{\theta_i}{2}\sin\frac{\theta_j}{2}e^{i(\phi_i-\phi_j)}d\phi_i\right.\nonumber\\
&&\left.-\frac{i}{2}\cos\frac{\theta_{ij}}{2}(d\chi_i+d\phi_i)\right)
\end{eqnarray}
and similar connection for the down-spinor from eq.(25),
\begin{eqnarray}
 \langle\downarrow_j|d|\downarrow_i\rangle
 &=&e^{i/2(\phi_i-\phi_j)}.  e^{i/2(\chi_i-\chi_j)} .\nonumber\\
&&\left(\frac{i}{2}\cos\frac{\theta_{ij}}{2}(d\chi_i+d\phi_i) -\right.\nonumber\\
&&\left.-i\sin\frac{\theta_i}{2}\sin\frac{\theta_j}{2}e^{i(\phi_j-\phi_i)}d\phi_i\right)
\end{eqnarray}
Geometrically in a frustrated spin system the parallel transport
of a spinor over a closed path on a sphere parameterized by
$\theta, \phi$ and $\chi$ implies open curve because the site at
the final point do not coincide with initial one. The Berry phase
for both frustrated up and down spinors in the spin glass system
will be obtained after integration over variation of $\phi$ by
$0\leq\phi \leq 2\pi$.
\begin{equation}
{\Gamma^\uparrow}_F= -2i\pi e^{i/2(\chi_j
-\chi_i)}\left(\cos\frac{\theta_{ij}}{2} -
\sin\frac{\theta_i}{2}\sin\frac{\theta_j}{2}\right)
\end{equation}
and
\begin{equation}
{\Gamma^\downarrow}_F= 2i \pi e^{i/2(\chi_i
-\chi_j)}\left(\cos\frac{\theta_{ij}}{2} -
\sin\frac{\theta_i}{2}\sin\frac{\theta_j}{2}\right)
\end{equation}
 These Berry phases ${\Gamma^\uparrow}_F$ or ${\Gamma^\downarrow}_F$ for frustrated system,
 are products of the helicity ($\chi$) dependent phase and the solid angle of
 spinor between the spinors based not only on the individual angles
 $\theta_i$ and $\theta_j$ of the spinors but also on
$\theta_{ij}$, the angle between the two.


In the absence of local frustration,
$\cos\frac{\theta_{ij}}{2}=0$, no spin conflict arises, indicating
the transport of spin vectors ideally parallel. As a result the
final site coincide with the initial leading to choose
$\chi_i=\chi_j$, $\phi_i=\phi_j$ and $\theta_i=\theta_j$ that
gives rise the phase for un-frustrated system.
\begin{equation}
\gamma^\uparrow=i\pi(1-\cos\theta_i)
\end{equation} and $$\gamma^\downarrow=-i\pi(1-\cos\theta_i)$$
These are the usual solid angles identical with eqs.(24) and (26)
respectively visualizing the Berry phase for the up/down spinor in
an isolated system. In a frustrated system,
$\cos\frac{\theta_{ij}}{2}=\pm 1$, acts as a signature of two
chirality that may act also an order parameter in the system.
 For an un-frustrated system even in the presence of a magnetic field which
is one of the very source of quantization, the helicity/internal
helicity depending Berry phase is not visible. This is only
realized in a frustrated spin glass system, where disorders offer
spin conflict to realize helicity depending phase along with the
solid angle.

At the end we would like to point out that the frustrated and
un-frustrated Berry phase could be a very source in developing the
nontrivial matrix Berry phase of two qubit state. In a very recent
communication [13] we have pointed out that due to frustration in
Quantum Hall system, the lowest Landau level LLL ($\nu=1$) is a
two qubit singlet state
\begin{eqnarray}
\Phi_1 (z) &=& \pmatrix{{u_i~~~~~u_j}\cr{v_i~~~~~v_j}}=(u_i v_j -
u_j v_i) \nonumber\\
&=& (u_i~~~~v_i)\pmatrix {{0~~~~~1}\cr{-1~~~~~0}}{u_j \choose v_j}
\end{eqnarray}
that has been identified as Hall qubit constructed from the
up-spinor $|\uparrow_i>={u_i \choose v_i}$. The non-abelian nature
of the connection on the Hall surface will remain if $i\neq j$.
\begin{eqnarray}
B_{\uparrow}&=& \pmatrix{{(u_i^*du_i +v_i^*dv_i)~~~~(u_i^*du_j
+v_i^*dv_j)}\cr{(u_j^*du_i +v_j^*dv_i)~~~~(u_i^*du_i +v_i^*dv_i)}}\nonumber\\
&=&\pmatrix{{\mu_i~~~\mu_{ij}}\cr{\mu_{ji}~~~~\mu_j}}
\end{eqnarray}
This is visualizing the spin conflict during the parallel
transport leading to non-abelian Berry phase. We realize in the
light of Hwang et.al [14] that non-abelian matrix Berry phase
created by frustration is responsible for the pumped charge flow
over a cycle by the singlet states in the Quantum Hall system.
\begin{equation}
{\gamma^H}_{\uparrow}=\pmatrix{{\gamma_i~~~~\Gamma_{ij}}\cr{\Gamma_{ji}~~~~\gamma_j}}
\end{equation}
Here $\gamma_i$ and $\gamma_j$ are the respective un-frustrated
Berry Phases for the ith and jth spinor as seen in eqs.(15) and
(17). $\Gamma_{ij}$ represents the off-diagonal Berry Phase
developed by the local frustration in the spin system. In the
absence of frustration there will be no development of matrix
Berry phase. Hence we would like to conclude that the matrix Berry
phase that is responsible for pumped charge to flow can be well
realized in the frustrated system.

{\bf Acknowledgements}
I like to acknowledge my home institute and specially The Abdus
Salam International Center for Theoretical Physics, Trieste,Italy
for giving me the full support for doing this work. Also I am
highly motivated by participating in the "School and Workshop on
Highly Frustrated Magnets and Strongly Correlated Systems: From
Non-Perturbative Approaches to Experiments" held during 30th July
to 17th August.07 at ICTP.

\pagebreak


\begin{thebibliography}{99}
\bibitem{paper1} G. Baskaran in {\it Physics of Disordered Solids}
 by Prabodh Shukla.\\P.W.Anderson and H. Hasegawa, Phys.Rev.{\bf 100}, 675 (1955).
\bibitem{paper2}G. Toulouse, Commun.Phys. {\bf 2}, 115 (1977).
 \bibitem{paper3} {\it Spin Glasses and Other
Frustrated Systems} by Debashish
 Chowdhury,World Scientific.
\bibitem{paper4} M.V.Berry, Proc.R.Soc.London {\bf A392},45(1984).
\bibitem{paper5} K. Ohgushi, S. Murakami and N. Nagaosa;
cond-mat/9912206.
\bibitem{paper6} D.Banerjee; Fort.der Physik {\bf 44} (1996) 323.
\bibitem{paper7} {\it Spin Glasses} by K.H.Fischer and J.A. Hertz,
Cambridge University Press.
\bibitem{paper8} B.Basu; J.Math.Phys. {\bf 34}, 737(1993).
\bibitem{paper9} P.Bandyopadhyay;Int. J.Mod.Phys. {\bf
A4},4449(1989).
\bibitem{paper10}D.Banerjee and P.Bandyopadhyay ; J.Math.Phys.{\bf 33}, 990 (1992).
\bibitem{paper11}D.Banerjee and P.Bandyopadhyay, Physica Scripta,
73,571(2006), ICTP preprint.
\bibitem{paper12}G.Baskaran and P.W.Anderson, Phys.Rev.{\bf B37}, 580 (1988).
\bibitem{paper13}D.Banerjee;"The Qubit rotation in QHE" communicated to
PRB.
\bibitem{paper14}N.Y.Hwang,S.C.Kim, P.S.Park and S.R.Eric Yang;arXiv;
cond. mat/0706.0947.
\end{thebibliography}
\end{document}